\documentclass[final,5p,times,twocolumn]{elsarticle}

\usepackage{hyperref}
\usepackage{nicefrac}

\journal{a journal: no}

\biboptions{authoryear}

\begin{document}

\begin{frontmatter}

\title{Modeling the price of Bitcoin with geometric fractional Brownian motion: a Monte Carlo approach}

\author{Mariusz Tarnopolski}
\address{Faculty of Physics, Astronomy and Applied Computer Science, Jagiellonian University, Krak\'ow, Poland}
\ead{mariusz.tarnopolski@uj.edu.pl}

\begin{abstract}
The long-term dependence of Bitcoin (BTC), manifesting itself through a Hurst exponent $H>0.5$, is exploited in order to predict future BTC/USD price. A Monte Carlo simulation with $10^4$ geometric fractional Brownian motion realisations is performed as extensions of historical data. The accuracy of statistical inferences is 10\%. The most probable Bitcoin price at the beginning of 2018 is 6358 USD.
\end{abstract}


\end{frontmatter}

\section{Introduction}\label{sect1}

Bitcoin, after being introduced by \citet{Nakamoto2008}, has been constantly gaining popularity. Its price and volume have increased by several orders of magnitude. Being the so-called cryptocurrency, i.e. an asset derived from mathematical cryptography, it is based on a new technology called the blockchain \citep{Bradbury2013,Ali2014}. Its other fundamental characteristics are: being decentralised, and having a fixed total number of coins: 21 million, with more than 16 million already in circulation.\footnote{\url{https://www.worldcoinindex.com/}, accessed on 2017-07-06.} Bitcoin is still very young, especially considering the fact that the last coin is to be mined around year 2140 \citep{Vranken2017}.

While Bitcoin does not have the features of a currency \citep{Baek2015,Dyhrberg2016} and is highly volatile \citep{Dwyer2015,Katsiampa2017}, it can serve as an effective diversifier \citep{Briere2015,Bouri2017} and be a complement to fiat currencies \citep{Carrick2016}. Moreover, it recently entered the phase of being weakly efficient \citep{Urquhart2016,Nadarajah2017} in the sense of the market efficiency hypothesis \citep{Mantegna2000}. On the other hand, Bitcoin market exhibits speculative bubbles \citep{Shiller2014,Cheah2015,Cheung2015}. Its high volatility and substantial speculative component are related to the fact that Bitcoin is still more a trading asset than a currency \citep{Dyhrberg2016,Katsiampa2017}.

The Bitcoin exchange rates have been already modeled and predicted by means of a noncausal autoregressive process \citep{Hencic2015}. On the other hand, financial time series, e.g. stocks, exchange rates, bonds or commodities, are characterised by the Hurst exponent \citep{Liu1999,Carbone2004,Matteo2005,Sanchez2008,Zunino2017}, and share the scaling properties with the fractional Brownian motion (fBm) \citep{Lo1991,Cheridito2003,Bayraktar2004,Cont2005,Gu2012,Areerak2014,Vukovic2015}. Therefore, in this work Bitcoin price forecasts are done via Monte Carlo simulations of a large sample ($10^4$) of geometric fBms as a stochastic process governing the price. The most probable prices are then derived from the empirical cumulative distribution function (CDF).

This paper is organised in the following manner. In Sect.~\ref{sect2} the data set and methods are described. Sect.~\ref{sect3} presents the results of the analysis, and in Sect.~\ref{sect4} discussion and concluding remarks are gathered.

\section{Data and method}\label{sect2}

\begin{figure*}[ht!]
\centering
\includegraphics[width=\textwidth]{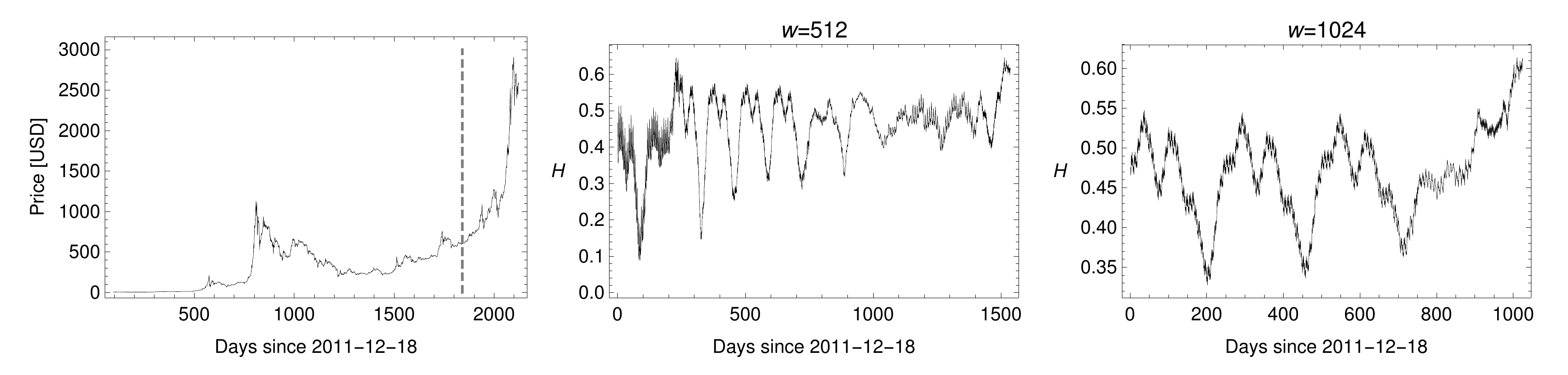}
\caption{{\bf Left panel:} Daily Bitcoin prices. The vertical dashed line marks the end of 2016. {\bf Middle panel:} Temporal evolution of $H$ with a sliding window of $w=512$. The horizontal axis denotes the starting date of a window; the last point in the plot refers to the last 512 days. {\bf Right panel:} The same as middle panel, but for $w=1024$.}
\label{plot1}
\end{figure*}

\paragraph{Data}
The examined data set spans from 2011-12-18 to 2017-07-06. The daily weighted prices were downloaded from BitcoinCharts.\footnote{\url{https://bitcoincharts.com/charts}, accessed on 2017-07-06.} A three-day gap from 2015-01-06 to 2015-01-08 was linearly interpolated based on prices on 2015-01-05 and 2015-01-09. The total number of data points is 2028. The data is displayed in Fig.~\ref{plot1}.

\paragraph{Hurst exponent}
To compute the Hurst exponent $H$, the wavelet approach using the Haar wavelet as a basis is employed \citep{Tarnopolski2016}; the error $\delta H$ is the standard error of the slope of the fitted linear dependence of the logarithmic variance of the wavelet coefficients' versus octave (see \citealt{Tarnopolski2016} and references therein for details). In order to make the size a power of 2, the examined data sets were padded left with zeros. For the whole dataset, $H=0.557\pm0.073$. To verify the validity of the employed method, the set of prices until the end of 2016 is examined. It is characterised by $H=0.495\pm0.102$ and consists of 1841 data points. The time dependence of $H$ for two sliding windows is displayed in Fig.~\ref{plot1}. An increase in the instantaneous value of $H$ could be observed recently.

\begin{figure*}[ht!]
\centering
\begin{tabular}{ccc}
\includegraphics[width=0.3\textwidth]{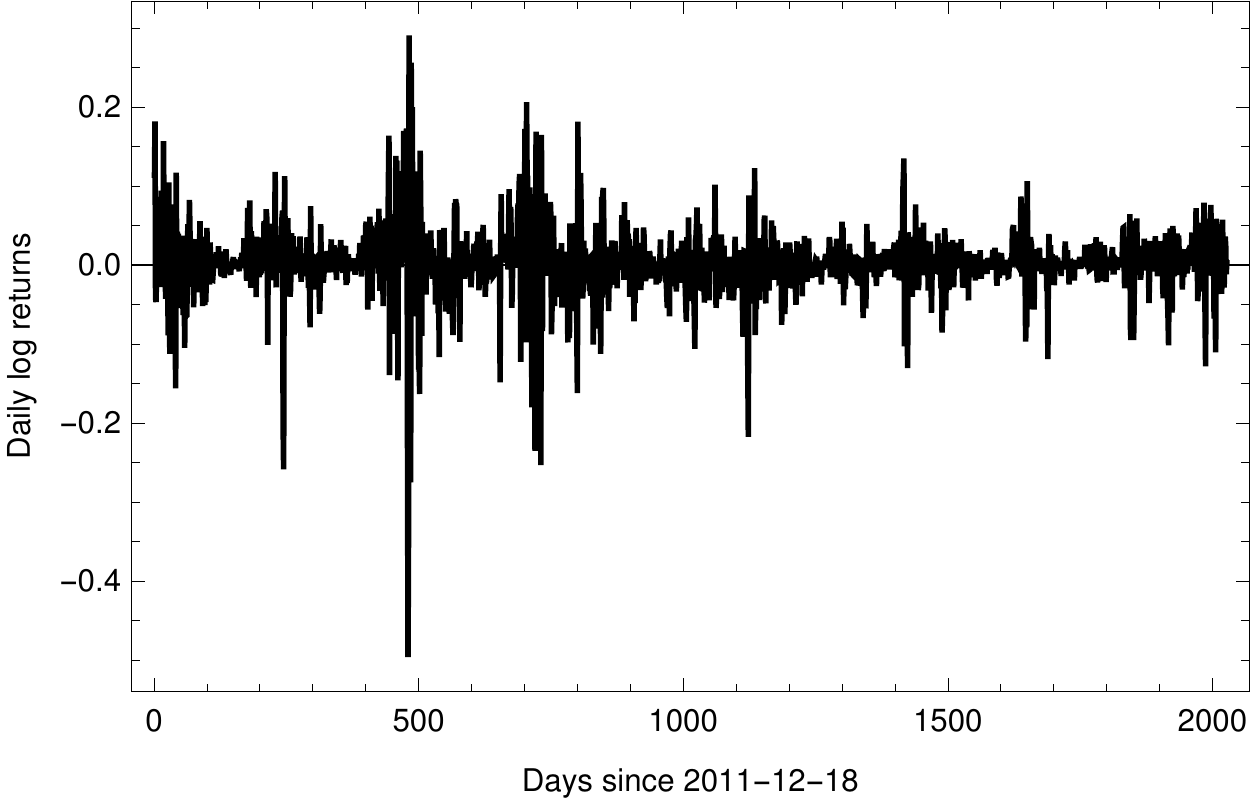}
&
\includegraphics[width=0.3\textwidth]{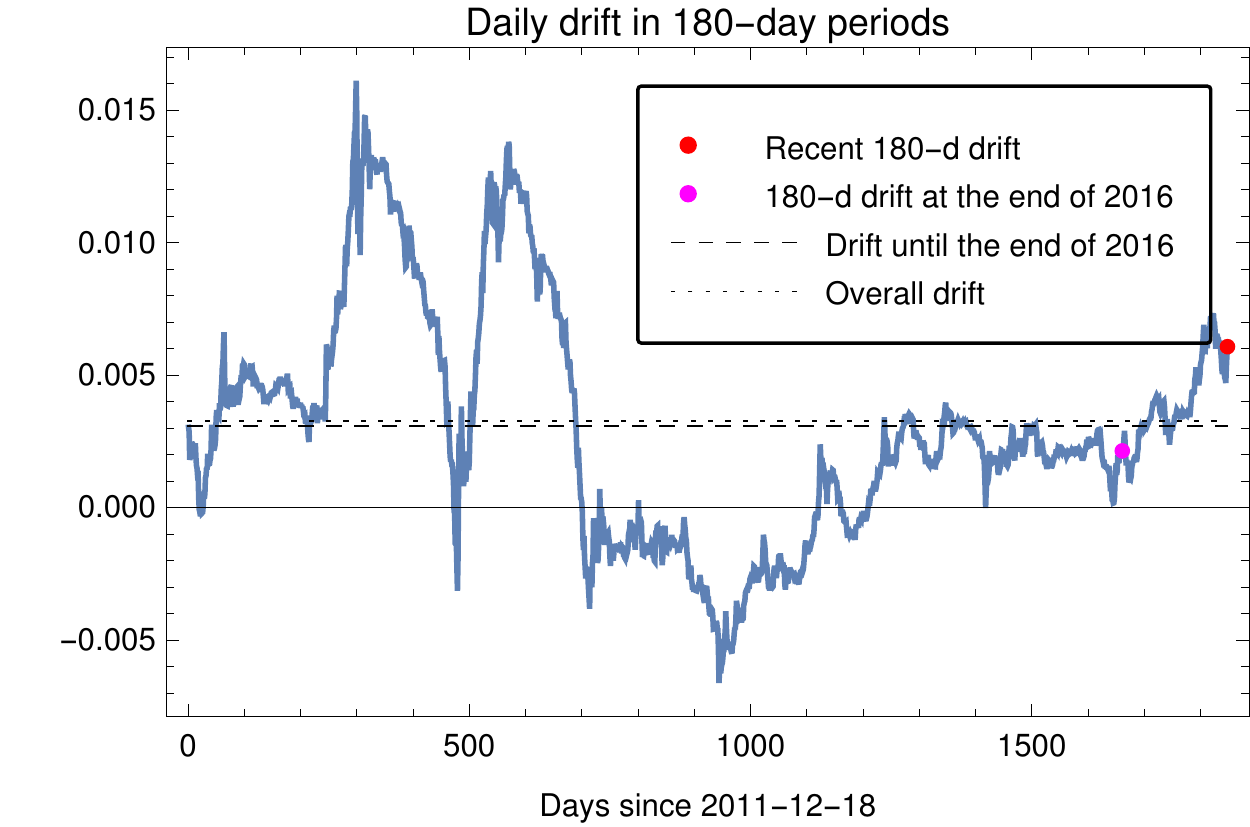}
&
\includegraphics[width=0.3\textwidth]{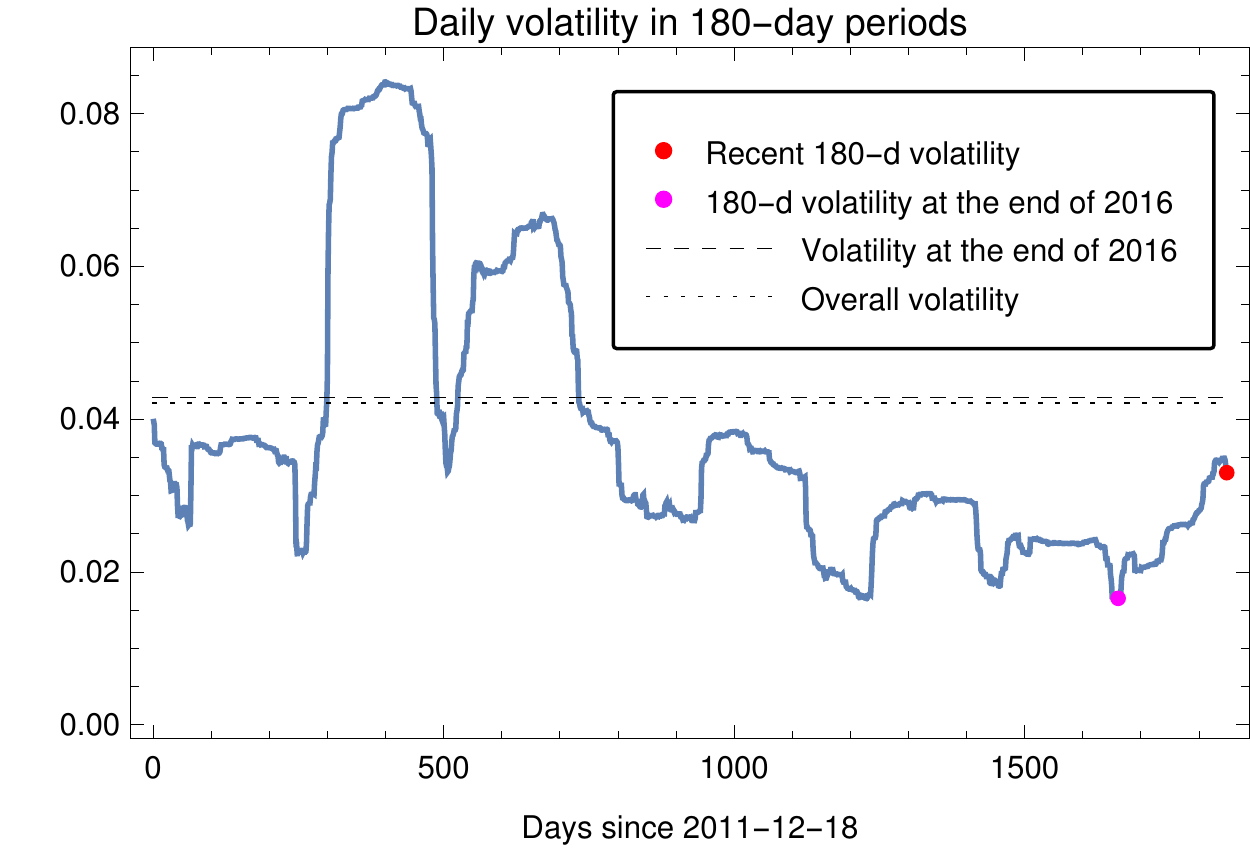}
\end{tabular}
\caption{{\bf Left panel:} Daily log returns. {\bf Middle and right panels:} Daily drift and volatility calculated over 180-day long periods. The horizontal axis denotes the starting date of a period (sliding window of length 180 days and advancing by one day). The red point marks the most recent values, the magenta ones stand for values as of the end of 2016, the dashed horizontal lines denote the values in the time interval from 2011-12-18 to the end of 2016, and the dotted lines mark the values obtained for the whole period from 2011-12-18 to 2017-07-06.}
\label{plot1.5}
\end{figure*}

To verify the approach from the next paragraph, the drift and volatility (taken as the mean and standard deviation, respectively, of the daily log returns; see Fig.~\ref{plot1.5}) are computed for daily log returns until the end of 2016 (yielding $\mu=0.0031$ and $\sigma=0.0428$), and for future forecasts the values appropriate for the whole historical data set are employed ($\mu=0.0033$ and $\sigma=0.0421$). The time dependencies of the drift and volatility of the whole period under examination are displayed in Fig.~\ref{plot1.5}. 

\paragraph{Methodology}
The idea exploited herein is to generate a large number ($10^4$) of fractional Brownian motion (fBm) realisations $B_t^H$ with Hurst exponent $H$, and insert them into the solution \citep{Dai1996,Biagini2008,Gu2012,Nguyen2013}
\begin{equation}
X(t)=X_0\exp\left(\mu t+\sigma B_t^H\right)
\label{eq1}
\end{equation}
of the stochastic differential equation describing a geometric fractional Brownian motion:
\begin{equation}
{\rm d}X(t)=\mu X(t){\rm d}t+\sigma X(t){\rm d}B_t^H,
\label{eq2}
\end{equation}
where $X_0=X(0)$ is the initial price of the extensions, taken as the last price in the historical data set, and $\mu$ and $\sigma$ are drift and volatility, respectively. The fBms are generated with $\mu$ and $\sigma$ the same as the historical prices yield: $B_t^H=B_t^H(\mu,\sigma)$. From the set of last prices of each realisation, an empirical CDF, $F_X(x)$, is constructed, and probabilities $P$ of reaching prices $X$ lower than $x$: $P(X\leq x)=F_X(x)$, or higher than $x$: $P(X>x)=1-F_X(x)$, are computed. The approach is first tested on a set of prices until the end of 2016, and then a prediction is made for the future 180 days. The computer algebra system {\sc Mathematica} is applied throughout this paper; in particular, fBms are generated as \texttt{FractionalBrownianMotionProcess[$\mu$,$\sigma$,H]}.

The distribution of $X(t)$ from Eq.~(\ref{eq1}) follows a log-normal distribution $\mathcal{LN}(m,s^2)$, with mean: $\exp\left(m+\frac{s^2}{2}\right)$, median: $\exp\left(m\right)$, and mode: $\exp\left(m-s^2\right)$, which in turn follow ${\rm mean} > {\rm mode} > {\rm median}$. Trivially, $F_X({\rm median})=\nicefrac{1}{2}$.

\section{Results}\label{sect3}

\begin{figure*}[ht!]
\centering
\includegraphics[width=\textwidth]{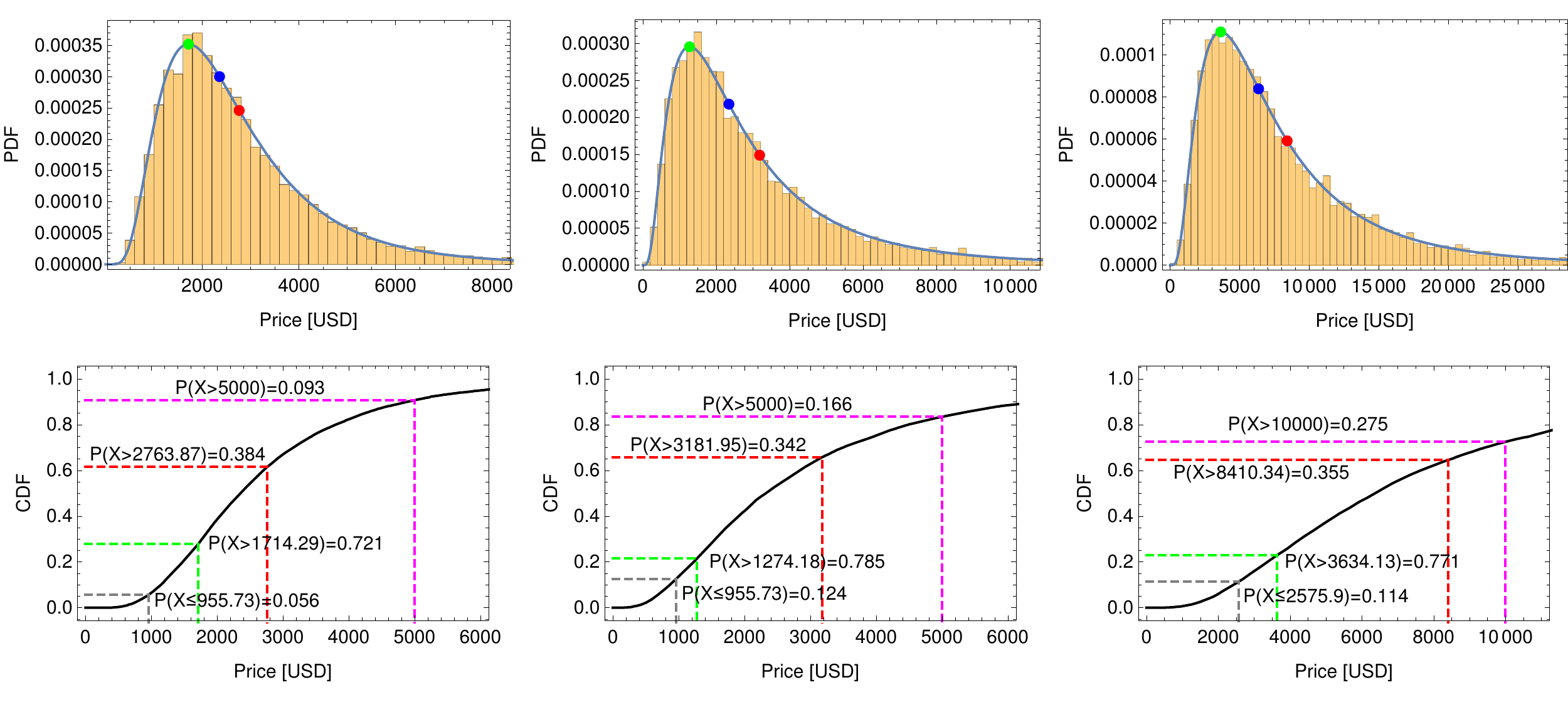}
\caption{{\bf Left column:} Predictions of the current Bitcoin price based on the historical data until the end of 2016 and its $H$ value of $0.495\pm0.102$. {\bf Middle column:} The same as te left column, but employing the current $H$ value of $0.557\pm0.073$. {\bf Right column:} Predictions of the price 180 days from the last historical value. {\bf Upper row:} Histograms displaying the empirical PDF overlaid with the best fit log-normal distribution. Highlighted are the mode (green dot), median (blue dot) and mean (red dot). {\bf Lower row:} The CDF with prices under consideration and their associated probabilities. Recall that $F_X(\rm median)=\nicefrac{1}{2}$.}
\label{plot2}
\end{figure*}

First, the set of prices until the end of 2016 is examined. It is characterised by $H=0.495\pm0.102$ and consists of 1841 data points. $10^4$ realisations of fBm (of length 187 each) are generated. Fig.~\ref{plot2} displays the results in its left column. The histogram is the empirical probability density function (PDF) of prices taken as the last value of the generated fBms (i.e., 2028 days after 2011-12-18, or on the 187th day of 2017), overlaid with a log-normal fit which yielded a mean of 2763.87 USD (red dot), and mode of 1714.29 USD (green dot). The probabilities of attaining a given price are calculated via the CDF; 955.73 USD is the price on 2016-12-31, and there was a 5.3\% chance of reaching a price smaller than this. Exceeding the mean 2763.87 USD had a 38.4\% probability, exceeding the mode 1714.29 USD had a 72.1\% chance of occuring, and a threshold of 5000 USD or more yielded a probability of 9.3\%. The median was at 2357.07 USD (blue dot).

The mean, mode and median for a log-normal distribution are different, as noted in Sect.~\ref{sect2}, but a log-transform of the data is normally distributed. For a normal distribution the the mean, mode and median are all equal to $m$, which denotes the most probable value, and through an inverse transform it turns into the median of a log-normal distribution, $\exp\left(m\right)$. Hence, the prediction of the Bitcoin price is taken as the median.

The middle column of Fig.~\ref{plot2} is the same as the left one, but with an updated $H=0.557\pm0.073$, i.e. the value for the whole historical data set. This is performed as the value of $H$ increased in the last months. Employing the actual $H$ yielded a mean price of 3181.95 USD, and a mode of 1274.18 USD. The probability of falling below 955.73 USD was 12.4\%, of exceeding the mean was 34.2\%, of exceeding the mode has a 78.5\%, and rising above the threshold of 5000 USD had 16.6\% chance of occurence. The median was at 2345.34 USD. Overall, the predicted price (i.e., the median) is about 10\% lower than the attained one, being a good agreement given the recent sudden rise in price, which lead to an increase of the instantaneous value of $H$ (compare with Fig.~\ref{plot1}).

To make predictions about the Bitcoin price 180 days from 2017-06-07, again $10^4$ fBm realisations were generated and inserted into Eq.~(\ref{eq1}). The results are displayed in the right column of Fig.~\ref{plot2}. The probability of falling below the current price of 2575.9 USD is 11.4\%, of exceeding the mean of 8410.34 USD is 35.5\%, of excedding the mode of 3634.13 USD is 77.1\%, and reaching a price higher than 10,000 USD has a chance of 27.5\%. The median is at 6358.32 USD, i.e. this is the predicted price to be attained in the beginning of 2018.

\section{Concluding remarks}\label{sect4}

The Bitcoin price was modeled as a geometric fBm \citep{Biagini2008}, and price predictions were put forward through a Monte Carlo approach with $10^4$ realisations. The predicted mid-2017 price, based on historical values until the end of 2016, taken as the median, was slightly---i.e., by about 10\%---underestimated. This is considered as a good agreement, thus justifying the applicability of the model. Therefore, price predictions for the beginning of 2018 were made in the same way. It is found, via the log-normal PDF and its CDF, that the price predicted as the median of a log-normally distributed set of realisations is 6358 USD. On the other hand, the chance of falling below the current price of 2575.9 USD is 11.4\%.

While the agreement of the predictions with reality is remarkable (i.e., considering the mean and median), it is well-known that large fluctuations in the market are more common than geometric fBm predicts. In fact, extreme events---like the 1987's Black Monday or the \mbox{2008-9} market crash---are virtually impossible under usual assumptions \citep{Jackwerth1996}. This is true for the classical \citet{Black1973} model as well. In case of Bitcoin, such an extreme event might be caused by the upcoming SegWit/UASF on 2017-08-01.\footnote{\url{http://www.uasf.co/}} A possible improvement might be to account for time variation of the volatility \citep{Areerak2014} and the Hurst exponent. On the other hand, other models that incorporate bullish and bearish parameters to account for big upward and downward jumps, e.g. \citep{Camara2008}, are perspective.

\paragraph{Disclaimer}
This manuscript is for information and illustrative purposes only. It is not, and should not be regarded as investment advice or as a recommendation regarding a course of action. The reader will make their own independent decision with respect to any course of action in connection herewith, as to whether such course of action is appropriate or proper based on their own judgment and their specific circumstances and objectives. The reader should seek a duly licensed professional for investment advice.

\section*{References}

\bibliography{mybibfile}

\end{document}